# Diffusion Stop-Layers for Superconducting Integrated Circuits and Qubits with Nb-based Josephson Junctions

Sergey K. Tolpygo, Denis Amparo, *Student Member, IEEE,* Richard T. Hunt, John A. Vivalda, and Daniel T. Yohannes

*Abstract*—**New technology for superconductor integrated circuits has been developed and is presented. It employs diffusion stop-layers (DSLs) to protect Josephson junctions (JJs) from interlayer migration of impurities, improve JJ critical current ($I_c$) targeting and reproducibility, eliminate aging, and eliminate pattern-dependent effects in $I_c$ and tunneling characteristics of Nb/Al/AlO$_x$/Nb junctions in integrated circuits. The latter effects were recently found in Nb-based JJs integrated into multilayered digital circuits. E.g., it was found that Josephson critical current density ($J_c$) may depend on the JJ's environment, on the type and size of metal layers making contact to niobium base (BE) and counter electrodes (CE) of the junction, and also change with time. Such $J_c$ variations within a circuit reduce circuit performance and yield, and restrict integration scale. This variability of JJs is explained as caused by hydrogen contamination of Nb layers during wafer processing, which changes the height and structural properties of AlO$_x$ tunnel barrier. Redistribution of hydrogen impurities between JJ electrodes and other circuit layers by diffusion along Nb wires and through contacts between layers causes long-term drift of $J_c$.**

**At least two DSLs are required to completely protect JJs from impurity diffusion effects – right below the junction BE and right above the junction CE. The simplest and the most technologically convenient DSLs we have found are thin (from ~ 3 nm to ~ 10 nm) layers of Al. They were deposited in-situ under the BE layer, thus forming an Al/Nb/Al/AlO$_x$/Nb penta-layer, and under the first wiring layer to junctions' CE, thus forming an Al/Nb wiring bi-layer. A significant improvement of $J_c$ uniformity on 150-mm wafer has also been obtained along with large improvements in $J_c$ targeting and run-to-run reproducibility.**

Manuscript received 1 August 2010. This work was supported in part by ONR under Grant N000140910079 and by the Office of the Director of National Intelligence (ODNI), Intelligence Advanced Research Projects Activity (IARPA), through the Army Research Office. All statements of fact, opinion or conclusions contained herein are those of the authors and should not be construed as representing the official views or policies of IARPA, the ODNI, or the U.S. Government.

S. K. Tolpygo, R.T. Hunt, J.A. Vivalda and D. Yohannes are with HYPRES, Inc., 175 Clearbrook Rd., Elmsford, NY 10523 (stolpygo@hypres.com; phone: 914-592-1190; fax: 914-347-2239).

D. Amparo is with the Department of Physics and Astronomy, Stony Brook University, Stony Brook, NY 11794-3800 (e-mail: denis.amparo@sunysb.edu).

S. K. Tolpygo is also with the Department of Physics and Astronomy and Department of Electrical and Computer Engineering, Stony Brook University, Stony Brook, NY 11794-3800.

## I. INTRODUCTION

SUPERCONDUCTING digital electronics is predominantly based on Rapid Single Flux Quantum (RSFQ) logic. Performance and operation margins of RSFQ circuits are very sensitive to circuit parameter spreads, especially to variations in the values of critical currents of Josephson junctions comprising the circuits. Therefore, the main requirement to high-yield manufacturing technology for superconducting digital circuits is to reproducibly deliver Josephson junctions having minimal deviations of critical currents, $I_c$, from the $I_c$s required by the circuit design and optimization. The most advanced fabrication technology capable of superconducting very large scale integrated (VLSI) circuits has been Nb-based technology that utilizes Nb/Al/AlO$_x$/Nb Josephson junctions and multiple layers of Nb for circuit inductors, interconnects, and signal routing.

It has been found recently that in Nb circuits, in addition to small and random variations of critical currents of JJs, there can exist large and systematic deviations of critical currents of JJs from the expected (design) values [1]-[4]. The former are caused by statistical fluctuations in the junction area and tunnel barrier transparency, and can be characterized by a standard deviation, $\sigma I_c$. The latter means that the value of $I_c$ deviation in a specific junction or a group of junctions, though varies from run to run, many times exceeds $\sigma I_c$, so the probability of this happening as a result of random fluctuations is statistically negligible. For instance, we have found that the $J_c$ in a JJ may depend on how the junction is connected to other circuit layers and on the area and shape of the contacting layers [1]-[2], whether the junction base electrode (BE) or counter electrode (CE) makes contact to Nb ground plane layer M0 [1]-[2], on the distance between the junction and the contact hole to other layers, and on the number of contact holes [3]. We have also found recently that the critical current density and the gap voltage increase over time in junctions stored at room temperatures if one or both of the junction electrodes are connected by Nb wire to Ti/Au or Ti/Pd/Au contact pads or just covered by a layer of Ti [4]. The effect is larger if CE is connected to a Ti-coated layer than if BE is connected.

The described above dependences of Josephson junction properties on the junction's environment and circuit patterns have been explained as resulting from hydrogen contamination of Nb circuit layers during wafer processing, with its subsequent migration towards or away from the $AlO_x$ tunnel barrier during the manufacturing cycle and later on upon its completion [3]-[6]. Long-term changes in $I_c$ were first found in Nb circuits with Pd coating [5],[6] and were suggested to be caused by hydrogen absorption and desorption.

It is well known that bulk Nb and Nb films can dissolve large amounts of hydrogen at room temperature, up to $c_H \sim 50$ atomic percent, where $c_H$ is the hydrogen content (H/Nb ratio). Hydrogen is the most mobile impurity [7]. Its diffusion coefficient in Nb at 300 K is $D \sim 10^{-5}$ cm$^2$/s; and diffusion activation energy is 0.106 eV, the lowest of all impurities (next is oxygen with the activation energy ~ 1 eV [8]).

Dissolved hydrogen changes many physical properties of Nb, e.g., it increases the lattice constant, resistivity, etc. (for a review, see [9]). Most importantly, it increases the work function of hydrogen-contaminated niobium, Nb(H), with respect to the clean Nb [10],[11]. As a result, the average height of the tunnel barrier in Nb/Al/$AlO_x$/Nb junctions and the barrier asymmetry become dependent on the $c_H$ in the CE near the tunnel barrier, as was proposed in our work [4] (see also [6]). As was also emphasized in our previous works, Nb base electrode in real Josephson junctions is coated by a thin Al layer which is only partially consumed by the oxidation forming $AlO_x$ tunnel barrier. As a result, the tunnel barrier in Nb/Al/$AlO_x$/Nb junctions is asymmetric (trapezoidal). The barrier height on the BE side, $\varphi_{BE}$, is determined by the work function of Al, and the barrier height on CE side, $\varphi_{CE}$, is determined by the work function of Nb, resulting in $\varphi_{CE} > \varphi_{BE.}$ [12]-[14]. Therefore, the presence of hydrogen in Nb base electrode has no effect on the tunnel barrier height and hence has much less effect on the critical current density of Josephson junctions than hydrogen dissolved in Nb counter electrode, as was explained in our work [4] (see also [6] and our work [15] in the present issue).

In addition to a reversible effect on Nb work function, it is possible that hydrogen can chemically react with the $AlO_x$ barrier and cause irreversible changes to its properties, e.g., create states with high transmission probability. Because diffusion of dissolved hydrogen in an integrated circuit occurs on a complex network of Nb wires interconnecting multiple junctions, resistors, and inductors, all with different diffusion coefficients and cross sections, complex concentration distributions may appear and depend on details of a particular circuit design. This can affect different circuits in a different, though always negative, and quite reproducible manner.

Obviously, in ideal technology hydrogen poisoning of Nb circuits should never happen. However, developing and maintaining this ideal technology can be very costly or it may not exist. Therefore, in this paper we propose and demonstrate a new technology in which Josephson junctions are protected from the effects of hydrogen (and other impurity) poisoning and interlayer migration by diffusion stop-layers.

In the following sections we discuss how and when hydrogen contamination of Nb layers of integrated circuits can happen during wafer processing. Then we describe our innovation and present experimental results demonstrating its success in eliminating pattern-dependent effects and variability of Josephson junctions as well as improving $J_c$ uniformity on 150-mm wafers and run-to-run reproducibility.

## II. Hydrogen Contamination of Nb Circuits: How, Where and When

Theoretically, $H_2$ concentration in air at atmospheric pressure is sufficient to saturate Nb with hydrogen at room temperature. The Nb surface presents a potential barrier for $H_2$ molecules and the native oxide on the surface works as a diffusion barrier, both prevent hydrogen absorption. However, in all situations when the surface oxide is removed hydrogen can easily dissolve in Nb. Hydrogen contamination can also occur because of the reaction with water molecules (in water, aqueous solutions, and moist air)

$$Nb + H_2O = NbO + H_2, \quad (1)$$

and due to a charge transfer process on a clean Nb surface

$$H_2O + e^- \rightarrow OH^- + H, \quad (2)$$

forming highly active atomic hydrogen which easily dissolves in Nb. Processes such as chemical etching, reactive ion (plasma) etching, chemical mechanical polishing (CMP), ion milling of Nb remove surface oxide and, hence, can produce hydrogen contamination.

During Nb film deposition by sputtering or other methods, hydrogen contamination can easily occur if there is sufficient residual hydrogen or water pressure in the vacuum chamber. After the deposition, hydrogen contamination is possible upon removing the deposited Nb film from the vacuum chamber because the clean surface of a freshly deposited film can readily react with air moisture.

In order to determine which integrated circuit fabrication steps result in hydrogen contamination of Nb layers in our process, we employed the following monitoring methods:

a) residual stress, σ, measurements of all metal layers after each process step;
b) electrical resistivity, ρ, measurements of all metal layers after each processing step;
c) optical emission spectroscopy (OES) of hydrogen lines during reactive ion etching (RIE) of all metal layers;
d) mass spectrometry during argon ion milling;
e) post-processing electrical characterization of test junctions and arrays of junctions with various connections to circuit layers, in the superconducting state at 4.2 K.

Absorption of hydrogen causes lattice expansion. Because Nb films are clamped to the substrate, absorption of hydrogen creates residual stress. For a clamped, (110)-textured Nb (as our films) the theoretical compressive biaxial stress increase is $\Delta\sigma/c_H = -9.6$ GPa. Somewhat lower values of $\Delta\sigma/c_H$ in the range from –6.5 GPa to –9 GPa were observed [9],[15], apparently due to a partial stress relaxation. Therefore, method a) allows for a nondestructive way of checking for hydrogen contamination and estimating its magnitude.

From numerous electrical measurements on bulk and thin film Nb, it is known that hydrogen impurities increase electrical resistivity of Nb with partial contribution $\Delta\rho/c_H$ in the range from 0.6 to 0.7 μΩ·cm per at.%. Therefore, a measured change in the resistivity of Nb layer after processing steps can be an indication of hydrogen absorption, and its magnitude can be estimated.

During RIE, all chemical elements containing in the layer being etch are being released into the RIE chamber and emit light due to excitation in rf plasma. By using an optical spectrograph coupled to the chamber we monitored the intensity of hydrogen optical emission lines at 486.1 nm and 656.5 nm to spot for abnormal increase in hydrogen emission, which would indicate hydrogen contamination of the layer.

TABLE 1 FABRICATION STEPS AND MONITORED PARAMETERS

| Step # | Processing Step Description | Monitored Parameters ρ | σ | OES | Changes? y/n |
|---|---|---|---|---|---|
| 1. | Nb ground plane deposition, layer M0 | + | + | | |
| 2. | M0 layer photolithography and RIE in $SF_6$ | | | + | no |
| 3. | $SiO_2$ interlayer dielectric deposition, layer I0 | | + | | no |
| 4. | I0 layer photolithography and RIE in $CHF_3$+$O_2$ mixture | | | + | |
| 5. | Nb/Al/$AlO_x$/Nb quad-layer formation by deposition and Al oxidation | + | + | | no |
| 6. | Counter electrode photolithography and RIE in $SF_6$ | | + | + | yes |
| 7. | Anodization in ammonium pentaborate in ethylene glycol solution, forms layer A1 | | + | | yes |
| 8. | A1 layer photolithography and Ar ion milling, forms layer M1 | + | + | | **yes** |
| 9. | Layer M1 (junction base electrode) photolithography and RIE in $SF_6$ | + | | + | yes |
| 10. | $SiO_2$ dielectric deposition, layer I1B-1 | | + | | no |
| 11. | Molybdenum resistor layer deposition, layer R2 | + | + | | no |
| 12. | Layer R2 photolithography and RIE in $SF_6$ | + | | + | no |
| 13. | $SiO_2$ dielectric deposition, layer I1B-2 | | + | | no |
| 14. | Layer I1B = (11B-1) + (I1B-2) photolithography and RIE in $CHF_3$+$O_2$ mixture, contact holes to BE and CE of JJs | | | + | |
| 15 | Nb wiring layer deposition, layer M2 | + | + | | yes |
| 16. | Layer M2 photolithography and RIE in $SF_6$ | + | | + | no |
| 17. | $SiO_2$ dielectric deposition, layer I2 | | + | | no |
| 18. | Layer I2 photolithography and RIE in $CHF_3$+$O_2$ mixture, contact holes to layer M2 | | | + | |
| 19. | Nb second wiring layer deposition, layer M3 | + | + | | yes |
| 20. | Layer M3 photolithography and RIE in $SF_6$ | + | | + | no |
| 21 | Ti/Au or Ti/Pd/Au chip contacts metallization deposition and pattering by lift-off, layer R3 | + | | | **yes** |

Monitored parameters are electric resistivity, ρ, residual films stress, σ, and the intensity of optical emission spectra (OES) of hydrogen. By bold "yes" shown steps after which substantial changes in the properties of circuit layers and/or Josephson junctions have been observed.

Our standard 11-layer fabrication process was described in detail in [16], [17]. Its cross-section is presented in Fig. 1 for a resistively-shunted JJ. In Table 1, we list the sequence of processing steps and show the parameters which were monitored (marked with + sign) in order to detect possible hydrogen contamination, and mark those processing steps after which substantial changes where detected either in the properties of the layers or in the tunneling characteristics of JJs.

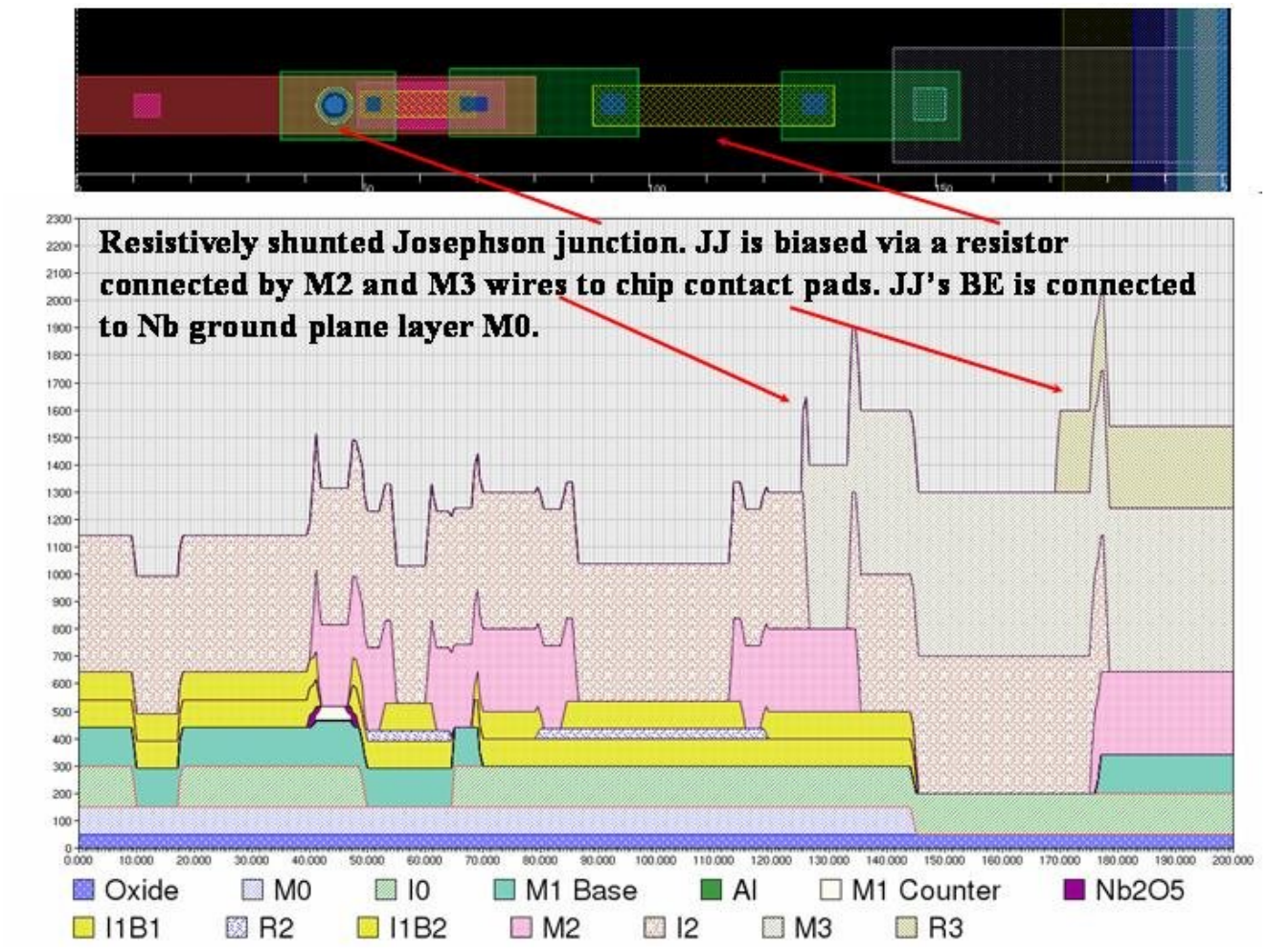

Fig. 1. Cross-section of our standard 11-layer fabrication process. Resistively-shunted JJ is shown. The junction is biased via resistor in R2 layer, connected to the chip contact pads (Ti/Au or Ti/Pd/Au) by Nb wiring in layers M2 and M3. The junction base electrode is connected to Nb ground plane layer M0 through a contact hole in the interlayer dielectric I0 (feature on the far left).

A small change in the residual stress was observed after counter electrode etching, which defines junctions, in comparison with the wafer state after Nb/Al/$AlO_x$/Nb quad-layer formation. No significant hydrogen optical emission intensity was observed during the RIE. Since JJs occupy a very small total area on the wafer, residual stress after etching represents mainly the stress in the base electrode plus Al/$AlO_x$ structure. Therefore, from the total stress in Nb/Al/$AlO_x$/Nb quad-layer and the observed change we can calculate individual stresses in the counter electrode and in the Nb/Al/$AlO_x$ base electrode structure. These data will be presented elsewhere.

By far the largest changes in the residual stress were observed after Ar ion milling of the ~ 36-nm-thick anodization layer on the surface of junction base electrode layer, as shown in Fig. 2. This anodization layer is formed all over the exposed wafer area right after CE etching and is needed to seal junctions' interior and form a protective layer of $Nb_2O_5$/$Al_2O_3$ around junctions' perimeter. Since only a small area of anodization around the junctions is needed (defined by A1 layer photolithography), the rest of the anodized area is removed using a neutralized beam of Ar ions (ion milling). This process is done in a cryopumped chamber with base pressure lower than $1\times10^{-7}$ Torr using a Kaufman source and rotating wafer for uniformity.

The change in the stress after anodization is consistently small and reproducible with the average increase of only 18.0

MPa and run-to run standard deviation, *sd*, of 2.1 MPa (11.7%), see Fig. 2. However, after removing this anodization layer by Ar milling, the stress in the remaining Nb film moves towards highly compressive, independently of the initial stress in the Nb/Al/$AlO_x$ multilayer. The average increase is -65.6 MPa and run-to-run variation is very large, *sd* = 47.7 MPa (73%). This clearly indicates that something happens to Nb films during Ar ion milling and that the amount of this something varies from run-to-run.

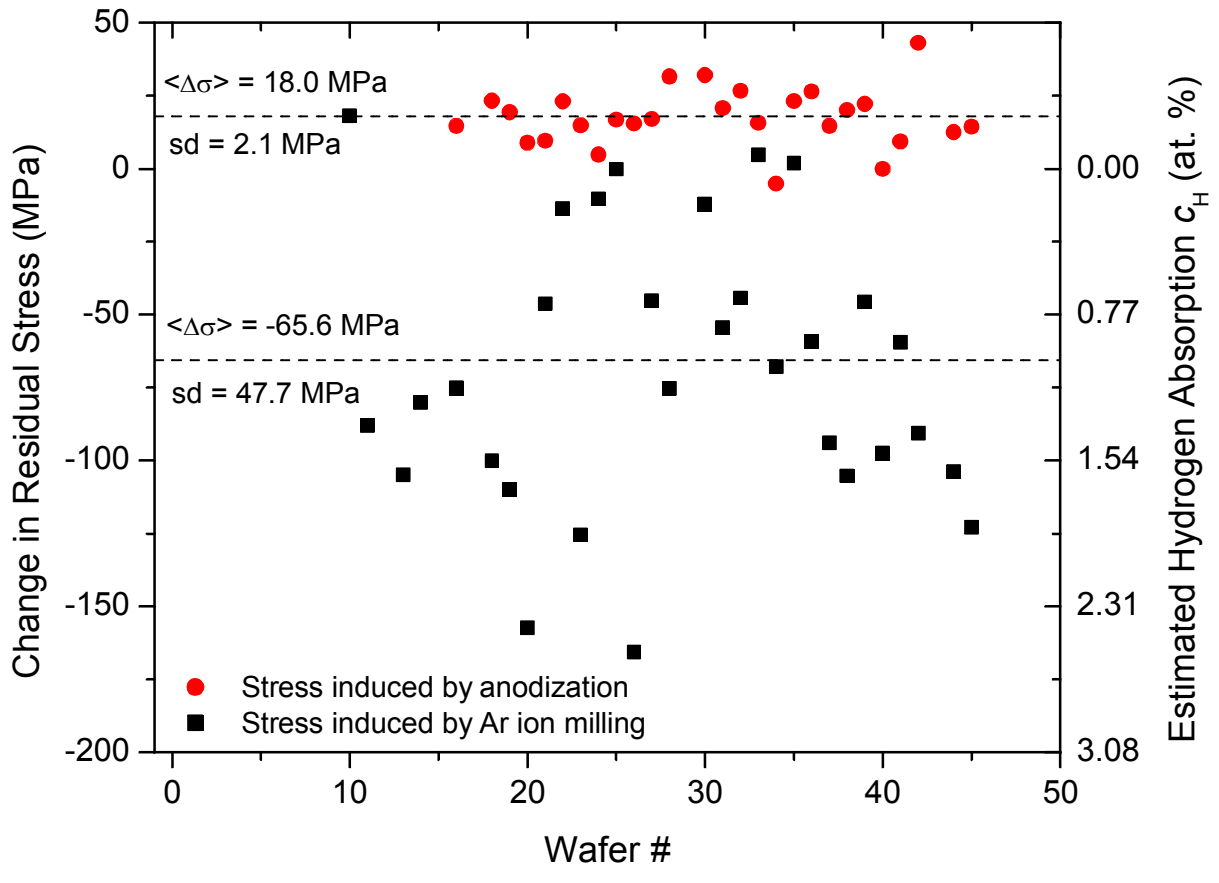

Fig. 2. Change in the residual stress in Nb/Al/AlOx (BE+barrier) structure after its anodization at 21 V (●), and after Ar ion milling that removes the anodized layer from ~ 95% of the wafer area, leaving it only around JJs (■). The thickness of the removed layer is ~ 40 nm.

We assume that the increase in compressive stress after Ar ion mill is the result of hydrogen absorption by Nb layers. Then, using the theoretical value of $\Delta\sigma/c_H$ = –96 MPa per at. % of H for perfectly clamped and (110)-textured films, we can estimate the average concentration of the absorbed hydrogen as $\langle c_H\rangle \approx 0.7$ atomic percent. If we use the experimental value for our films $\Delta\sigma/c_H$ = – 65 MPa per at.% [15], we get a somewhat larger average value $\langle c_H\rangle \approx 1.0$ at.% and run-to-run variation of ±0.7 atomic percent. The source of this hydrogen is not clear. It can be dissolved in the $Nb_2O_5/Al_2O_3$ layer during anodization in the water-containing electrolyte and then released and re-absorbed or knocked into Nb, or it can come from $H_2O$ and hydrocarbons adsorbed on the wafer surface. Hydrogen poisoning of Si during Ar ion milling is well known in semiconductor industry, although its source has never been completely understood [18], [19].

A noticeable increase in the resistivity of the Nb layer remained after milling has also been observed in comparison with the anodized Nb/Al/AlOx multilayer before Ar ion milling. If we assign all the observed increase in ρ to hydrogen contamination (the anodized layer does not contribute to the conductivity), we get $c_H$ = 3±1 atomic percent. This is a factor of 3 larger than the estimate from the compressive stress increase. It is likely that only a part of the ρ increase observed is due to hydrogen contamination, whereas another part is due to some disordering of Nb film by Ar atoms. More work is needed in order to quantify H concentration in Nb layers.

Patterning of the junctions' base electrode unavoidably involves removing Al/$AlO_x$ layer (anodized or not) in order to etch Nb. It is done differently in different technologies and in different labs. Most frequently, wet etching in acidic solutions is used [5],[20],[21]. Hydrogen contamination of Nb then occurs, as was observed in [5], with the typical concentration of the absorbed hydrogen of 1 to 2 atomic percent, similar to what we see after ion milling. Strong changes of Nb properties in Nb/Al/$AlO_x$ layers were also observed after RIE in $SF_6$ [20]. However, possible role of hydrogen was not considered at that time.

Also, a big change in the critical current density was observed in the junctions connected to Ti/Au or Ti/Pd/Au chip contact pads by Nb wires, no matter which Nb layer (M2 or M3) was used. This effect was studied in detail in [4], [15], and was explained as resulting from diffusion of hydrogen from JJ's counter electrode into Ti.

## III. Diffusion Stop-Layers for Integrated Circuits

Examining the process cross-section in Fig. 1, we see that there are two major diffusion passes which may affect JJs and create pattern-dependent changes in $J_c$ – from BE to contacting layers and from CE to contacting layers. Impurity diffusion effects on JJs can be substantially reduced or eliminated if we can prevent or impede diffusion of impurities to and from junction electrodes. This can be done by using a diffusion barrier (stop-layer) – a material with much smaller diffusion coefficient and impurity solubility than the host material it protects. For protecting Nb layers from hydrogen diffusion, there is a big choice of metals and dielectrics satisfying these requirements. However, we need to maintain a good superconducting contact between Nb layers in the contact holes and vias. This eliminates all dielectric materials from consideration. Among metals, we can use only those which would superconduct by proximity with Nb and provide high critical current for contacts holes with minimum size used in the process, e.g., $I_c$ > 2 mA (the higher the better). Also, they should be compatible with the existing deposition systems. From these considerations we have chosen Al and Mo, also because they are used in the process anyway.

Al has extremely low solubility of hydrogen, ~ $6\times10^{-6}$ at. % at room temperatures. Hydrogen diffusivity in Al at room temperatures can vary by a few orders of magnitude according to different data, being in the range from $10^{-10}$ $cm^2/s$ to $10^{-7}$ $cm^2/s$ [22]-[24]. Even in the worst case, it is at least two orders of magnitude lower than in Nb. Combined with the extremely low solubility, this makes Al a very efficient diffusion barrier for hydrogen.

We placed the first DSL between junctions' base electrode and Nb ground plane to prevent in- and out-diffusion to/from the BE. The second DSL was placed between junctions' counter electrode and Nb wiring layer M2 to prevent in- and out-diffusion of hydrogen to/from CE. The third DSL was placed between wiring layers M2 and M3. The third DSL serves for extra protection of JJs and for protecting Nb inductors which are made mainly in M2 layer: changes in impurity concentration in Nb inductors may change magnetic field penetration depth and thus sheet inductance, which would have a negative effect on circuit performance.

All DSLs were deposited in-situ with the corresponding Nb layer on the surface of the previously processed layers. Ar sputter etching was used to remove native oxide from the surface of the bottom Nb layers in the contact holes and vias. The first (aluminum) DSL was deposited in-situ with the JJ layer, which now was deposited as Al/Nb/Al/$AlO_x$/Nb penta-layer. The second DSL was deposited in-situ with the first wiring layer, thus forming Al/Nb or Mo/Nb bi-layer. In the same manner, the third DSL was deposited in-situ with the second wiring layer M3. The cross-section of the new process with three DSLs is shown in Fig. 3.

The minimum thickness of DSLs was determined by observing the reduction in pattern-dependent effects in $J_c$ of JJs with various connections to circuit layers. For Al it was found to be ~ 3 nm and ~ 5 nm for Mo. The maximum thickness is determined by the desired level of the critical current of Nb/DSL/Nb contacts. We set this level at ~ 30 mA for circular contacts of 2-μm diameter. This gave ~ 10 nm upper limited for Al DSL thickness and less for Mo. From these considerations, the thickness was chosen to be ≈5 nm for both materials.

The best results were obtained with aluminum DSLs which completely eliminated effects of BE connection to M0 layer studied in [1]-[3] and effects of CE to Ti/Au contact pad connections studied in [4]-[6],[15]. With molybdenum DSLs we found a substantial reduction in the value of the second effect but it was not completely eliminated, apparently due to a higher diffusivity and solubility of H in molybdenum than in aluminum.

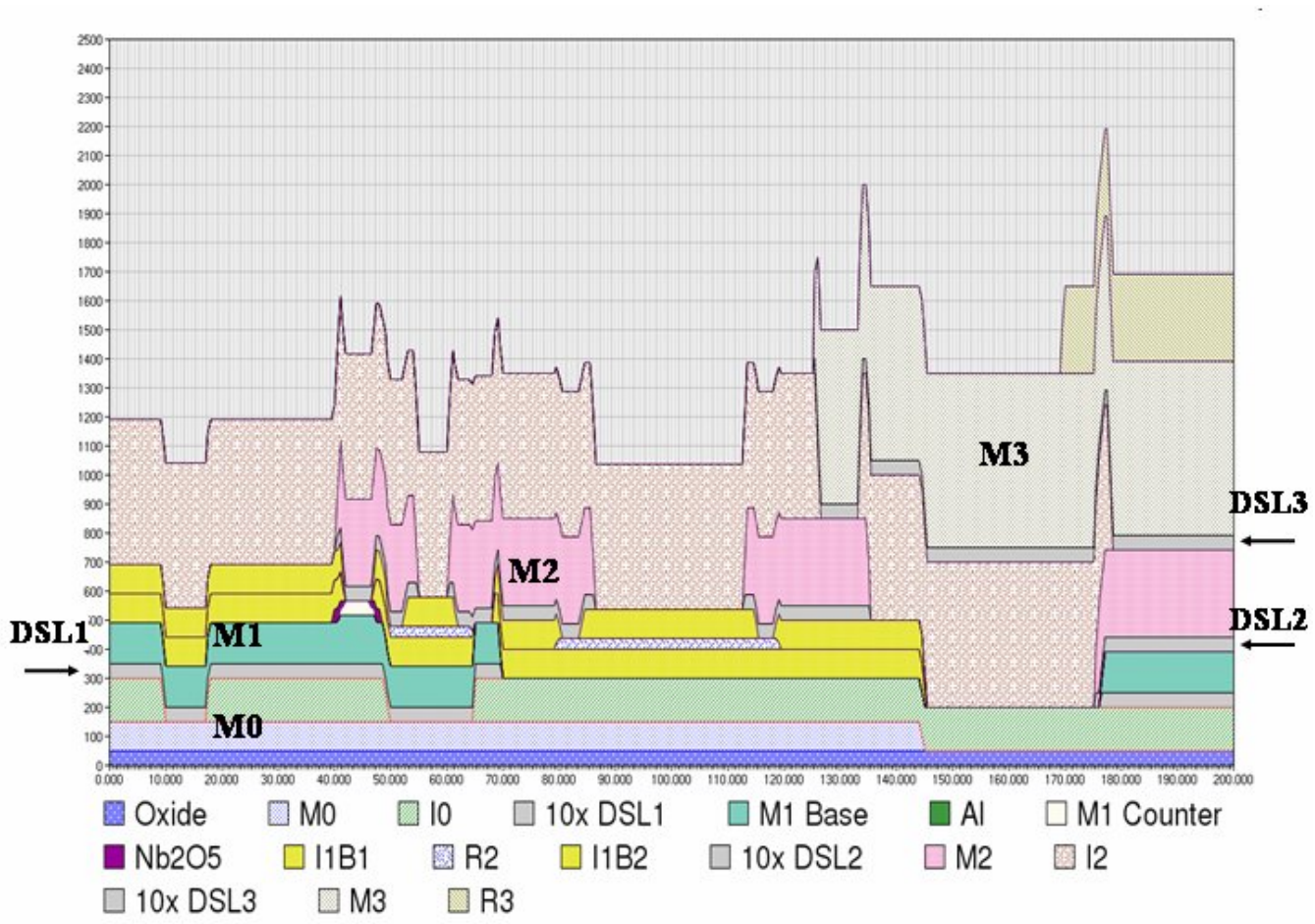

Fig. 3. Cross-section of the new process employing three diffusion stop layers: under BE, under the first wiring layer, and under the second wiring layer. Resistively-shunted junction is shown as in Fig. 1. All superconducting contacts between Nb layers have now structure Nb/DSL/Nb, e.g. Nb/Al/Nb. All junctions are connected to other layers as Nb/DSL/JJ/DSL/Nb. The DSLs are shown 10 times thicker than their actual thickness to be visible in this scale.

The described DSL method can be applied to any process layer structure and also to planarized processes with CMP, and to processes with any number of Nb layers. Just each pair of Nb layers should be separated by a DSL. The minimum number of required DSLs is twice the number of JJ layers (making it two in the present technology).

It is possible to deposit both DSLs protecting JJ in-situ with the JJ layer as DSL/Nb/Al/$AlO_x$/Nb/DSL multilayer. However, we found the processing in this case to be somewhat more difficult than when the top DSL is deposited separately, under the next wiring layer. Similarly, a less convenient way was found to be the process when the first DSL is deposited in-situ on top of Nb layer below the junction layer (M0 ground plane in our case).

Each DSL was patterned along with the Nb layer underneath which it is located, using the same photoresist mask (otherwise it would short the circuit). Aluminum DSLs were either etched in tetramethylammonium hydroxide solution in DI water or dry etched in $Cl_2/BCl_3$ mixture, or Ar ion milled. No difference was found in the results. Molybdenum DSLs were dry etched in $SF_6$ plasma in-situ right after Nb etching.

## IV. Experimental Results

Fig. 4 shows the typical effect of junction BE connection to Nb ground plane (layer M0) through a contact hole in I0 layer. The cross-section of this type of connection was shown in Fig. 1. This result is very similar to the previously presented [1]-[4]. The critical currents of two junctions deviate significantly from all other junctions – the grounded JJ and its nearest neighbor because they are the closest to the contact hole in I0.

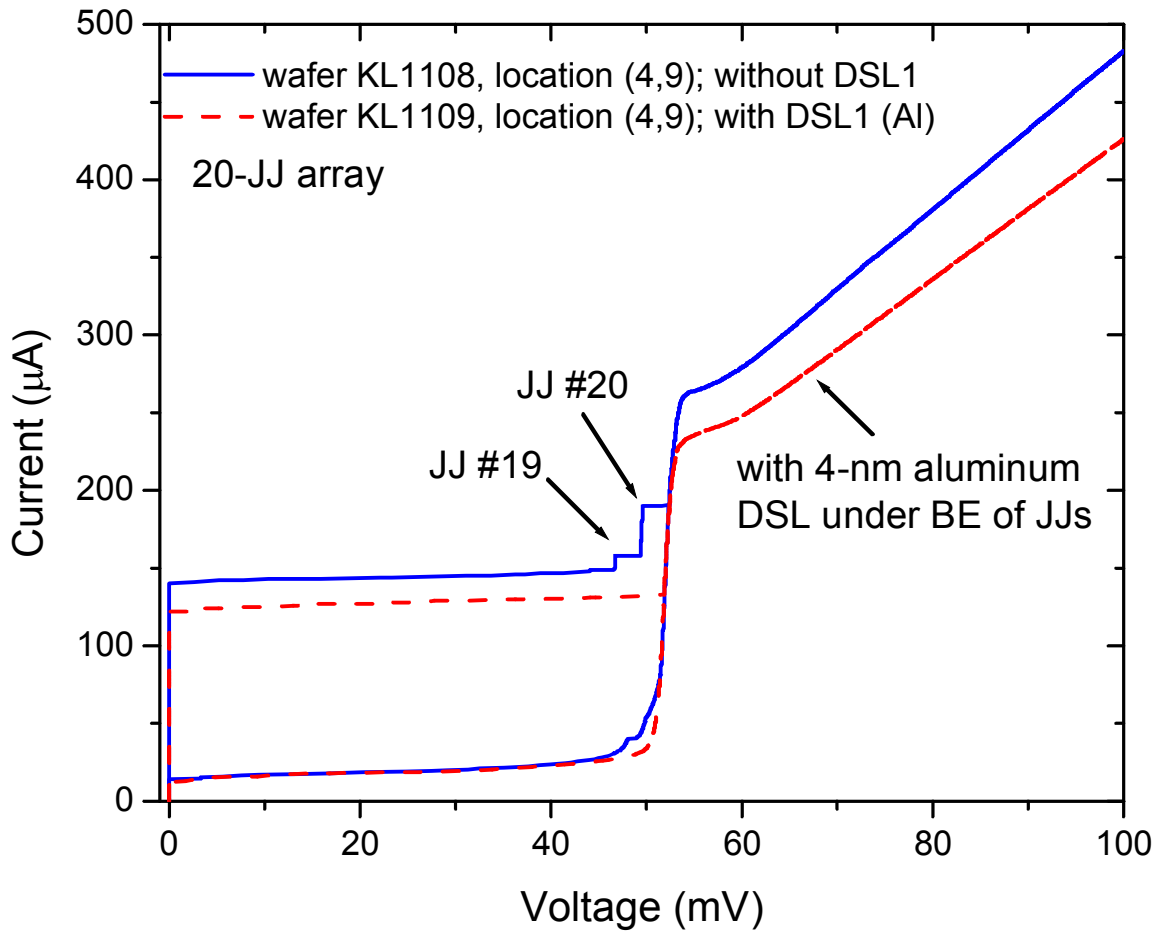

Fig. 4. Current-voltage characteristics of a series array of 20 identical junctions with the last JJ (#20) connected by its BE to Nb ground plane through a contact hole. Two junctions (#20 and #19) closest to this I0 contact hole change their critical current significantly during the processing as a result of this proximity (solid line). No changes in the critical current at any distance to the contact hole was observed if a DSL (4-nm layer of Al) was placed underneath the BE (dash line), as indicated by uniform switching of all 20 JJs.

We have fabricated and studied over 40 wafers with Al diffusion stop-layer underneath BE (deposited in-situ as described in Sec. III). We have not seen any effects of JJ connection to the ground plane on these wafers. That is, by adding this DSL the problem described in [1]-[4] has been completely solved. As an example, in Fig. 5 we show the data for 17 wafers (with 4-nm tick aluminum DSL1 and without it). As a measure of the effect we use the relative deviation of the $I_c$ of the junction with BE connection to the circuit ground plane from the next highest critical current in the array, i.e. of junction #20 from junction #19 in Fig. 4.

Similarly, the second DSL (between CE and M2 layer) almost completely removed the effect of CE connection to Ti-coated chip contact pads on JJs and associated long-term drift of $J_c$ in junctions with this connection. By now we have fabricated close to 30 wafers with this second DSL, and in only a few observed a small effect associated with hydrogen migration into the Ti layer of contact pads. It is possible that in these few wafers the surface roughness was such that 5-nm thick Al layer did not cover the surface of junctions' CE completely or had some pinholes. In this case, the diffusion of hydrogen was impeded but not completely prevented. Therefore, a slightly thicker DSL might be needed. This will be further investigated.

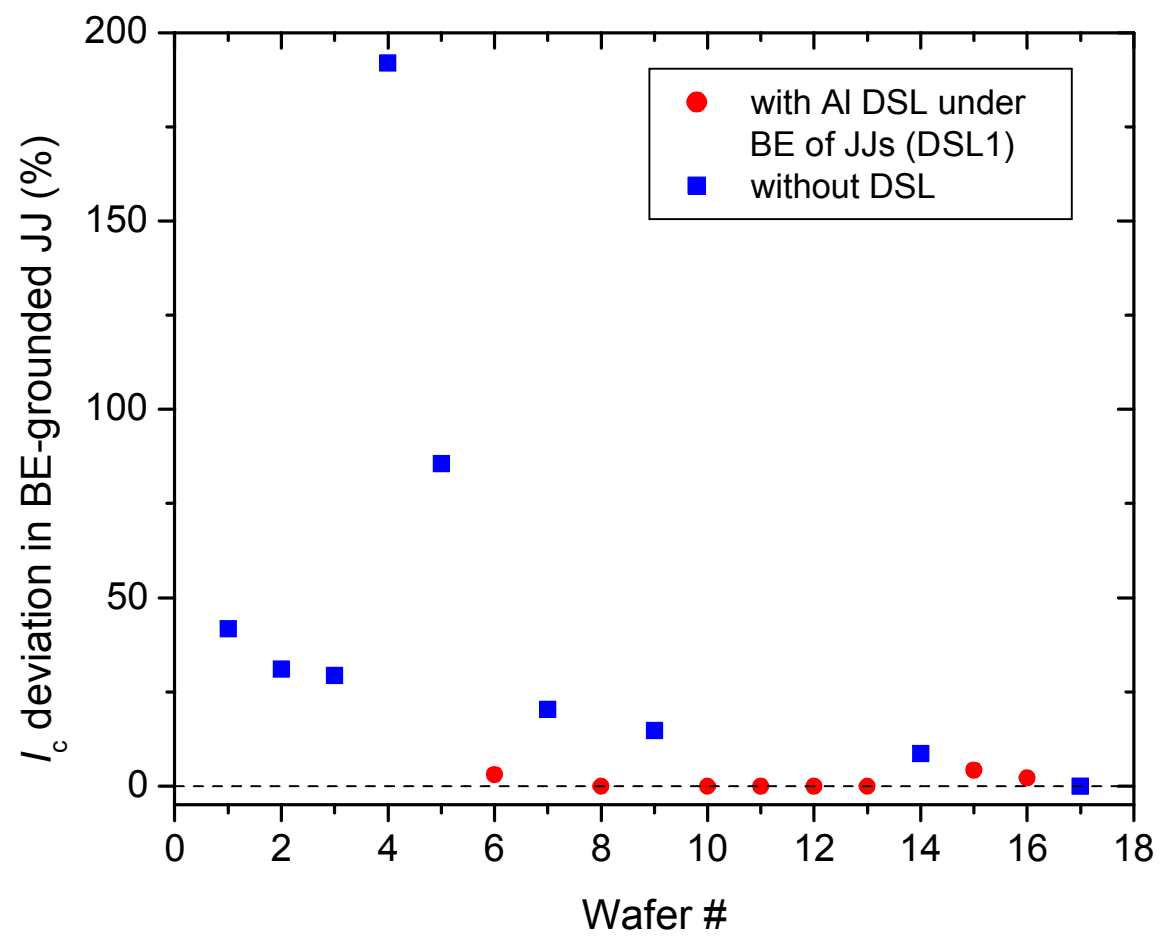

Fig. 5. Relative deviation of the critical current of junctions with BE connected to the circuit ground plane from otherwise identical junctions without this connection on wafers with a 4-nm Al diffusion stop layer (●) under junctions' base electrode (DSL1 in Fig. 3) and without this DSL (■).

Overall, the addition of DSLs has been found to have a very positive effect also on $J_c$ targeting and run-to-run reproducibility as shown in Fig. 6 – Fig. 8. The critical current density was measured in five fixed locations on each wafer, using the same test structure. These locations are the center of the wafer (0,0) and centers of each of the four quadrants (+,+), (–,+), (–,–), (+,–). It can be seen that on many parts of the wafers the $J_c$ targeting improved substantially, and run-to-run reproducibility improved by almost a factor of 10 (the standard deviation decreased by almost tenfold).

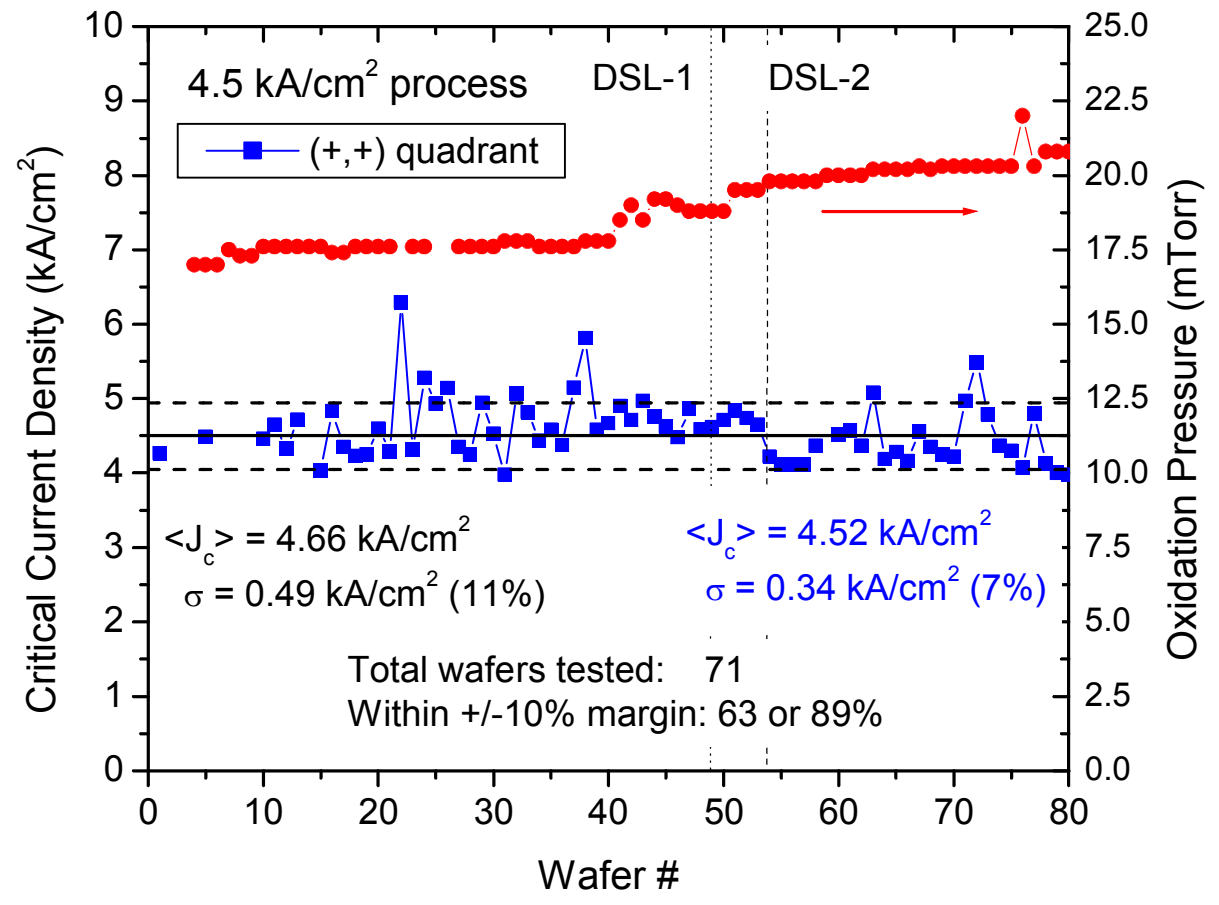

Fig. 6. Run-to-run reproducibility of $J_c$ in the first (+,+) quadrant of 150-mm wafers for 4.5-kA/cm$^2$ process. The targeted margin imposed by the circuit design requirements is ±10% shown as horizontal dash lines. Two vertical dash lines indicate when the first DSL (under BE) and the second DSL (between CE and Nb wiring layer M2) were introduced. The average critical current and its standard deviation are shown before and after the DSLs introduction.

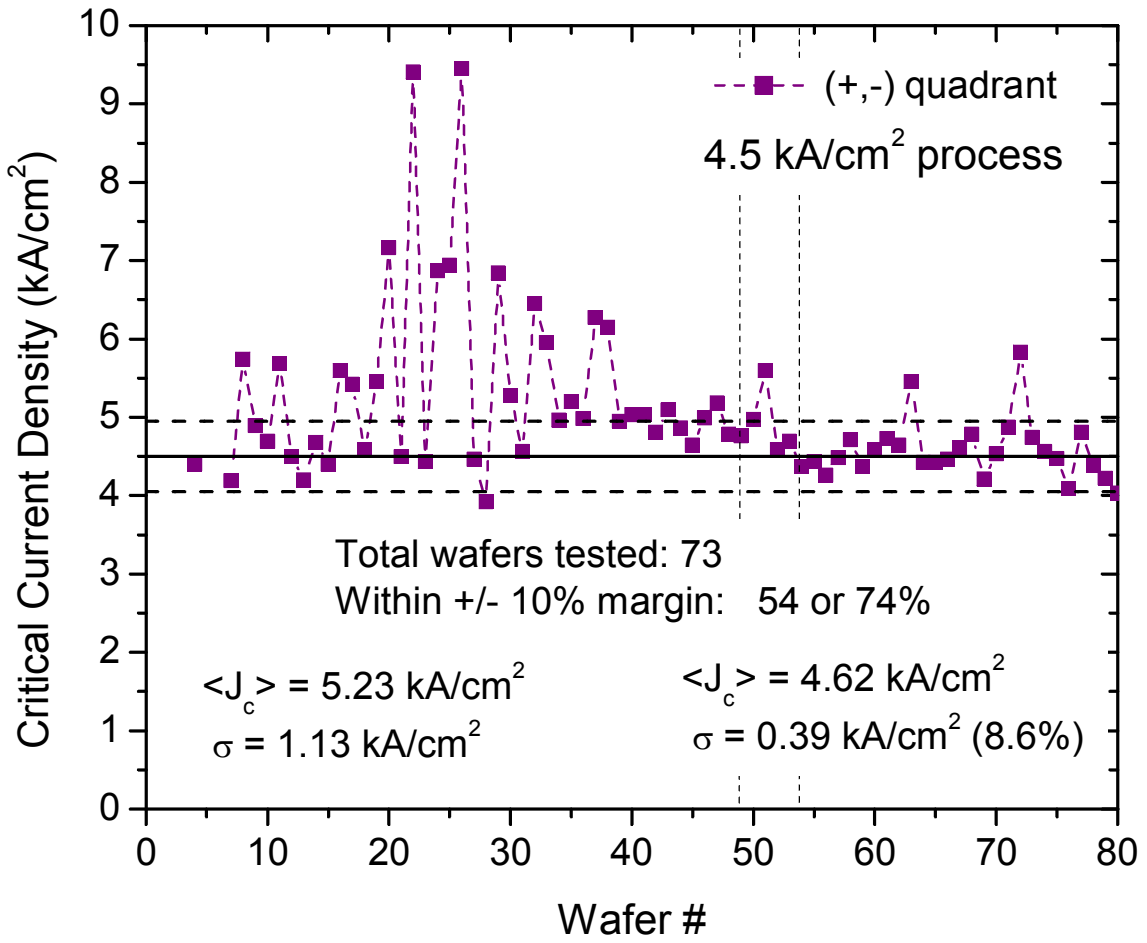

Fig. 7. Run-to-run reproducibility of $J_c$ in the fourth (+,-) quadrant of 150-mm wafers for 4.5-kA/cm$^2$ process. All notations are the same as in Fig. 6. The $J_c$ targeting and reproducibility (standard deviation σ) improved significantly after the introduction of the DSL layers.

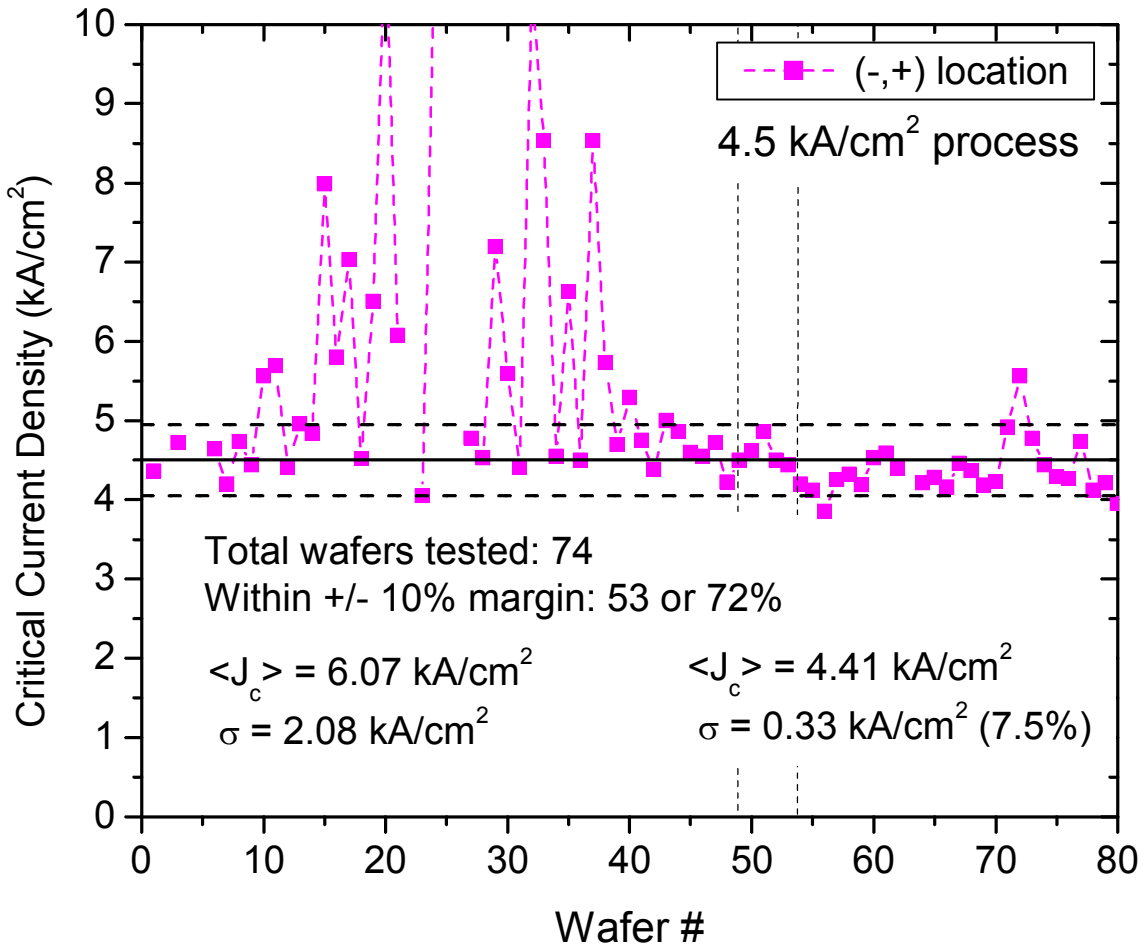

Fig. 8. Run-to-run reproducibility of $J_c$ in the second (-,+) quadrant of 150-mm wafers for 4.5-kA/cm$^2$ process. All notations are the same as in Fig. 6. The $J_c$ targeting and reproducibility dramatically improved (more than 6-fold for standard deviation σ) after the introduction of the DSL layers (especially of the first, under BE).

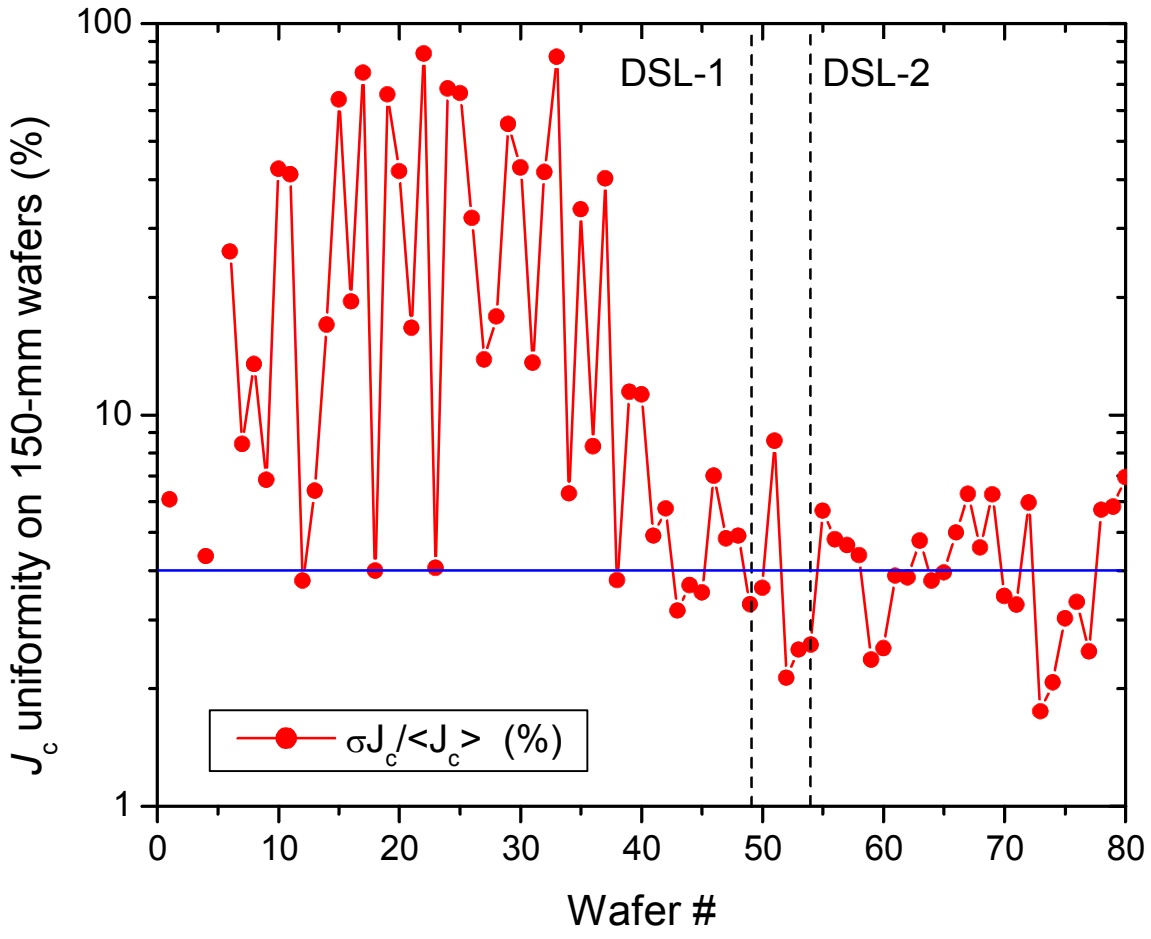

Fig. 9. Uniformity of the critical current density, $\sigma J_c/\langle J_c\rangle$, on 150-mm wafers. Here $\sigma J_c$ is the standard deviation, $\langle J_c\rangle$ is the average $J_c$. Significant improvement of uniformity was achieved by adding the DSL layers.

The addition of the DSLs also resulted in the overall improvement in $J_c$ uniformity across the wafers as shown in Fig. 9.

## V. Conclusion

In conclusion, we have developed a new technology for Nb integrated circuits that employs diffusion stop-layers between contacting Nb layers in order to prevent interlayer migration of impurities and their effect on tunnel barrier. These DSLs were deposited in-situ along with Nb Josephson junctions and wiring layers. Complete elimination of circuit pattern-dependent and contacting-layer-dependent effects on $J_c$ of JJ was found with this technology. These effects were associated with interlayer migration of hydrogen impurities and its effect on the tunnel barrier height. Long-term drift of $J_c$ related to diffusion of hydrogen impurities from junctions' BE into Ti and (or) Pd layers of contact pads can of course be eliminated by simply removing Ti and Pd layers from the contact metallization stack. In cases when Ti and Pd layers are needed (e.g., for wafer bumping for flip chip bonding) their effect can be eliminated by placing a DSL under the stack (between the uppermost Nb layer and contact metallization).

The described DSL approach can be easily extended to planarized processes with CMP and to processes with any number of superconducting and junction layers. We also applied it to superconducting qubits with Nb junctions to protect their surface from oxidation and/or formation of two-level systems (TLS) on their surface and at the interfaces between different layers, and with the substrate. As a DSL material we extensively studied Al and Mo, and found Al to work better. Different materials may be required if the Josephson junction material is not Nb.

## Acknowledgment

We would like to thank Saad Sarwana for testing numerous process diagnostic structures and Alex Kirichenko for designing them. Many discussions with Deborah Van Vechten, Vasili Semenov, and Timur Filippov are gratefully acknowledged. We also thank Anubhav Sahu and Andrei Talalaevskii for reporting to us all instances of abnormal critical currents of Josephson junctions in superconducting digital integrated circuits. Dave Donnelly, as always, provided excellent equipment support.